\documentclass{article}
\usepackage[T1]{fontenc}
\usepackage[utf8]{inputenc}
\usepackage{ismir} % Remove the "submission" option for camera-ready version
\usepackage{amsmath,cite,url}
\usepackage{graphicx}
\usepackage{color}
\usepackage{bm}
\usepackage{booktabs}
\usepackage{amssymb}
\usepackage{subcaption}
\usepackage{comment}
\usepackage{float} 

\title{Do Music Foundation Models \\Embed Pitch in Helical Structure?}

\multauthor
  {Hayato Yagi$^{*,1}$ \hspace{1cm}
   Shinnosuke Takamichi$^{*,1,2}$ \hspace{1cm}
   Rin Sato$^3$}
  {{\bf Keitaro Tanaka$^3$ \hspace{1cm}
   Shigeo Morishima$^3$}\\
   $^1$ Keio University, Japan \hspace{0.5cm}
   $^2$ The University of Tokyo, Japan \hspace{0.5cm}
   $^3$ Waseda University, Japan\\
   {\tt\small $^*$\{hayatobuti523,shinnosuke\_takamichi\}@keio.jp}
  }
\def\authorname{H. Yagi, S. Takamichi, R. Sato, K. Tanaka, and S. Morishima}

\begin{document}
\setlength{\abovedisplayskip}{3pt} % 数式の上のマージン
\setlength{\belowdisplayskip}{3pt} % 数式の下のマージン

\maketitle

\begin{abstract}
    This study analyzes the intermediate representations of music foundation models (MFMs) and reports the geometric structures used to represent pitch information. By inputting isolated musical notes into trained MFMs and analyzing their principal components, we reveal that the representations form a helical structure reflecting the octave periodicity of pitch. Furthermore, we show that the clarity and geometry of this helical structure vary not only across models but also with the acoustic properties of the input signals. Our analysis provides a novel approach for clarifying the internal mechanisms of MFMs.
\end{abstract}

\section{Introduction}\label{sec:introduction}
    % 最後に少しだけ行をはみ出すのをやめたい
Foundation models have recently advanced music information processing, following their success in natural language processing (NLP). In particular, self-supervised learning (SSL) on large-scale music data has produced powerful music-specific models, known as music foundation models (MFMs)~\cite{jukebox, musicgen, musiclm, mert, sonido}.
MFMs exhibit strong versatility in both music generation and understanding tasks such as classification and analysis, drawing attention for their potential to support new forms of music creation~\cite{jukemir, sonido}.

Despite their impressive capabilities, the internal mechanisms of MFMs remain unclear. In particular, how they acquire and use musical knowledge such as pitch and harmony is not well understood. To address this, recent studies have analyzed intermediate representations learned by MFMs, using probing methods to implicitly evaluate what musical knowledge is encoded~\cite{Syntheory,symbolic_concepts}. Trained on human-composed music, MFMs may encode such knowledge as latent structures.
However, it still remains unclear how such knowledge is explicitly represented as a geometric structure. Clarifying this would provide important insights into whether MFMs understand musical knowledge as structured representations.
% In NLP, large language models are known to represent periodic concepts as geometric structures~\cite{month, mod}, but similar analyses of MFMs are limited.

\begin{figure}[t]
  \centering
  \includegraphics[width=0.98\linewidth]{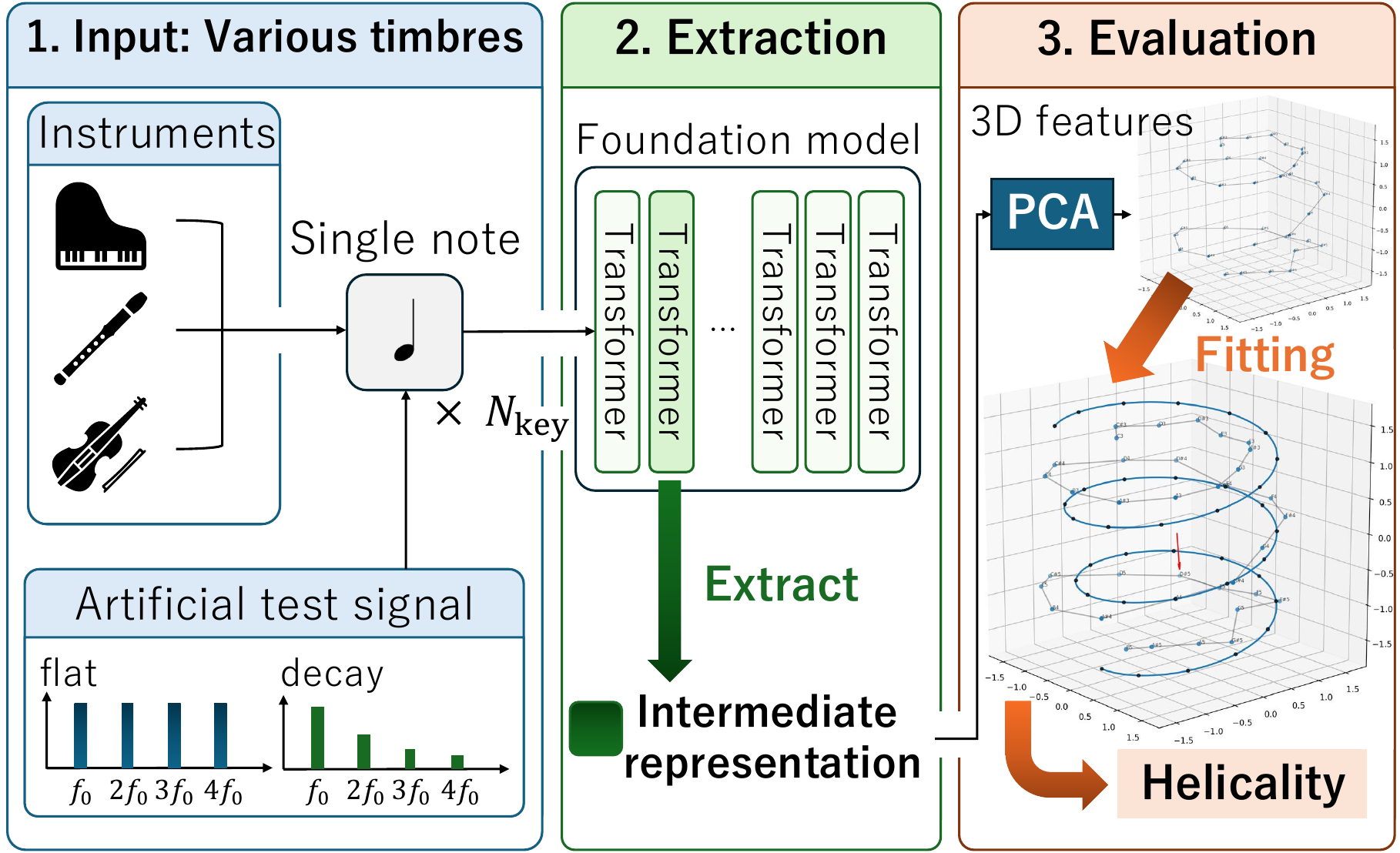}
  \vspace{-3mm}
  \caption{Overview. We extract pitch-dependent features from a music foundation model and evaluate their helical structure using PCA and parametric fitting.}
  % \caption{Overview of this study. We analyze whether pitch representations in a music foundation model exhibit a helical structure.}
  \label{fig:concept}
\end{figure}

As a first step toward understanding the internal geometry of MFMs, we focus on pitch, a fundamental musical element. In music psychology, pitch has long been modeled as a helical structure combining pitch height and octave periodicity~\cite{pitch_helix}. We hypothesize that MFMs may exhibit this structure and investigate whether their intermediate representations form a pitch helix. We extract pitch-related features from the representations and quantify the degree to which they follow a helical structure using a dedicated metric (Fig.~\ref{fig:concept}). 
% Furthermore, we analyze how this structure emerges and varies across different model architectures, as well as how they are influenced by acoustic properties of the input signals, including timbre and harmonic structure. This work offers new insights into MFM internals, aiming to improve transparency and controllability. 
In this study, we analyze two generative MFMs and show that (1) the helical structure emerges across a wide range of instrument sounds, and (2) this structure is primarily driven by specific harmonics, as revealed by experiments with artificial test signals.
Our code is available at \url{https://github.com/takamichi-lab/mfm-pitch-helix}.

\section{Related Work}\label{sec:related-work}
    
\subsection{Interpretability of Foundation Models}\label{sec:FM_analysis}
Understanding the intermediate representations of foundation models has become a key research topic, especially in NLP. Prior work shows that models like GPT-2~\cite{gpt-2} encode periodic concepts (e.g., days, months) as circular geometric patterns~\cite{month}, suggesting they form human-like conceptual structures beyond simple statistical patterns~\cite{mod, time, marjieh2025numberlargelanguagemodel, zhu-etal-2025-language}.

In music information processing, probing~\cite{probing} has been widely used to assess whether intermediate representations encode musical attributes by measuring how well a linear classifier predicts them.
It has been applied to MFMs such as Jukebox~\cite{jukebox} and MERT~\cite{mert}, demonstrating that intermediate layers encode musical attributes such as pitch, chords, tempo, and genres~\cite{Syntheory,mert,symbolic_concepts,2025universalmusic,sonido}. 
Some of the studies~\cite{Syntheory,symbolic_concepts} also report that deeper layers better encode pitch and chords.
However, these studies primarily assess what musical attributes are encoded in the intermediate representations via probe performance. Thus, unlike in NLP, research that directly examines how music-theoretical concepts are represented as geometric structures in MFMs remains limited.

\subsection{Pitch Helix}\label{sec:pitch_helix}
\begin{figure}[t]
  \centering
  \includegraphics[height=3.5cm]{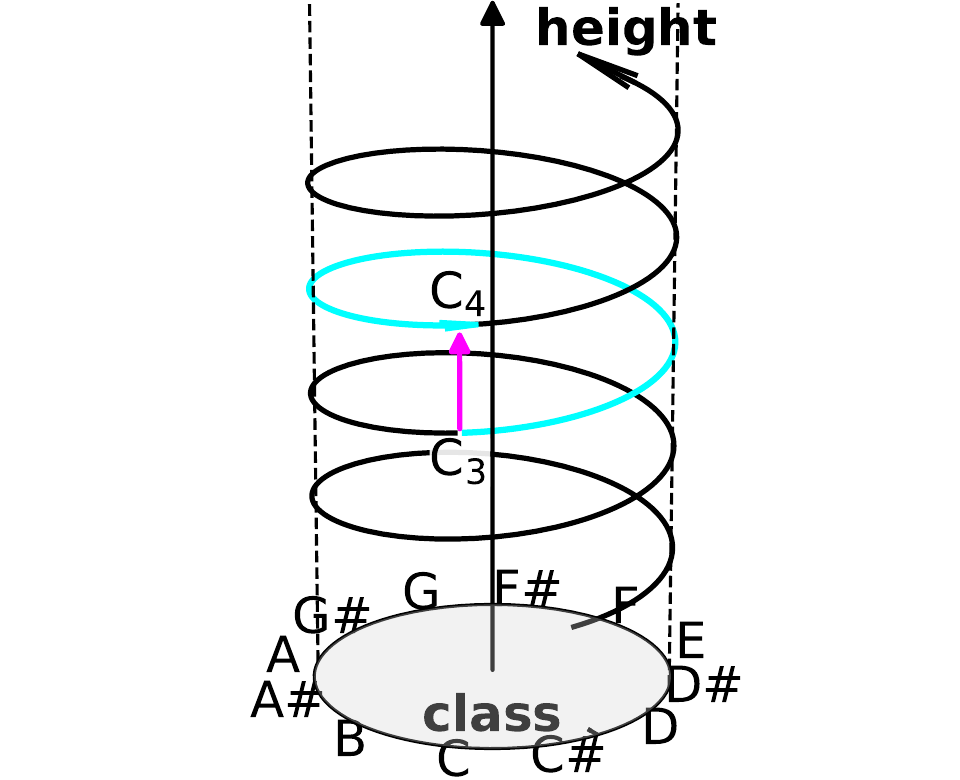}
  \vspace{-3mm}
  \caption{Shepard’s pitch helix~\cite{pitch_helix}. Vertical height and circular angle correspond to pitch height and pitch class, respectively.}
  \label{fig:pitch_helix}
\end{figure}

In music psychology, pitch is conceptualized along two perceptual dimensions: pitch height, corresponding to absolute frequency, and pitch class, representing note identity within the twelve-tone octave cycle (e.g., \texttt{C}, \texttt{C$\sharp$}, \texttt{D}), regardless of octave. This dual representation leads to the principle of octave equivalence, where tones that are one octave apart are perceived as similar. 
To capture this property, the pitch helix has been proposed as a model that arranges pitches in a three-dimensional spiral~\cite{pitch_helix}. In this structure, vertical position reflects pitch height, while the angular position represents pitch class, resulting in tones recurring at the same angular position in every octave (Fig.~\ref{fig:pitch_helix}).

% Recent studies have shown that such helical structures can emerge from raw music signals without supervision~\cite{cqt_helix}.
% Although this method reveals the helical structure of raw music signals, it lacks quantitative evaluation. 
Recent work shows that such helical structures can emerge from raw music signals without supervision~\cite{cqt_helix}. However, it lacks quantitative evaluation.
To overcome this, a metric called \textit{Helicality} has been proposed to measure how closely three-dimensional points fit to an ideal helix~\cite{helicality}. In our study, we apply the Helicality to the intermediate representations of MFMs in order to determine whether they encode pitch in a helical form.

\subsection{Timbre and Harmonic Structure}\label{sec:timbre}

Instrument sounds are complex tones composed of a fundamental frequency $f_0$ and multiple harmonics. The frequencies, amplitudes, and temporal evolution of these components determine timbre.
The harmonic structure depends on the sound production mechanism and physical properties of the vibrating body, and varies across instrument types such as string, wind, and percussion~\cite{physics_inst}.

\begin{figure}[t]
  \centering
  \includegraphics[width=0.90\linewidth]{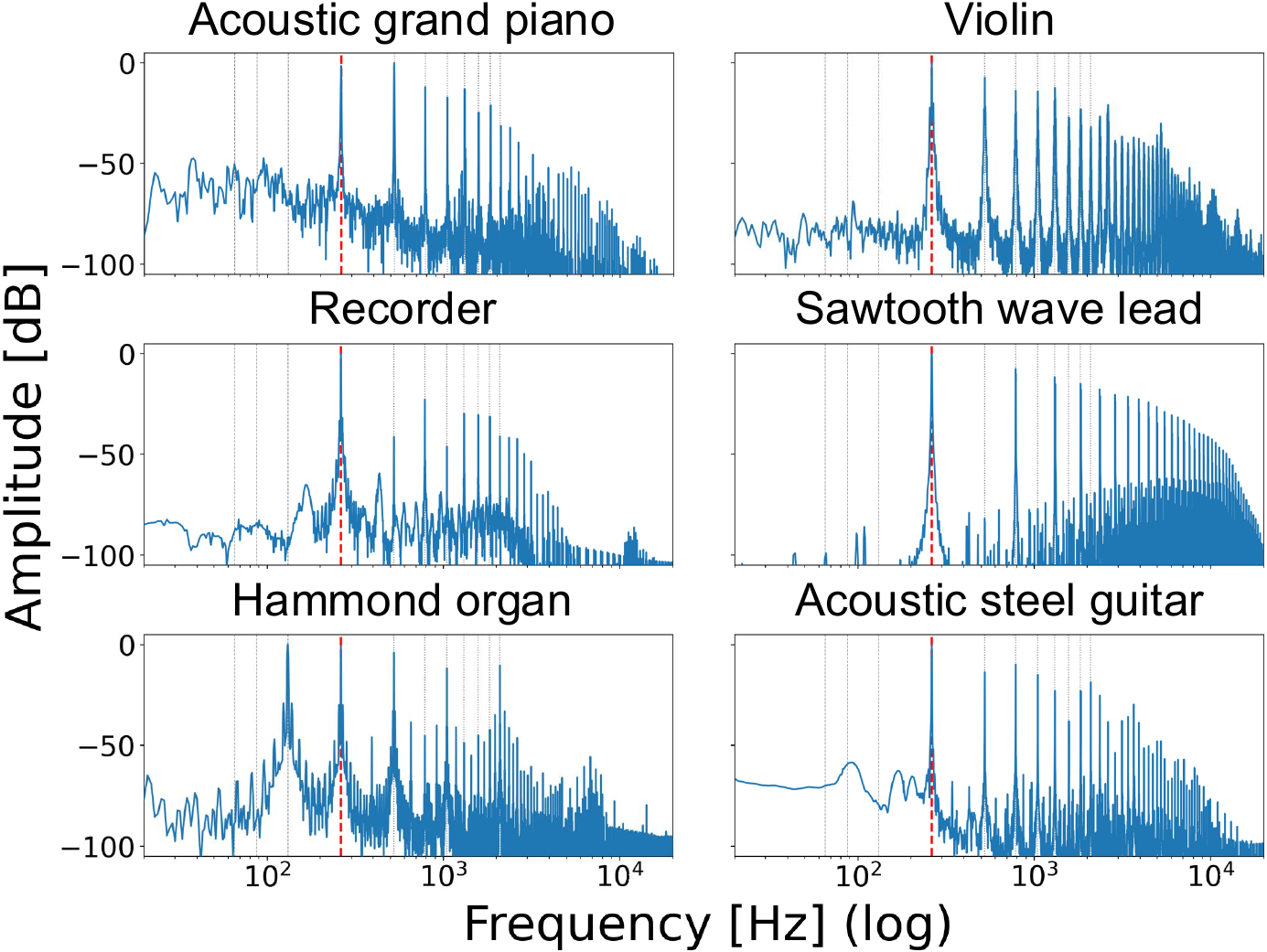}
  \vspace{-3mm}
  \caption{Amplitude spectra for various instruments. The red dashed line indicates the fundamental frequency $f_0$, and the black dashed lines indicate subharmonics $f_0/h$ or harmonics $h f_0$, where $h$ is an integer.}
  \label{fig:instrument_fft_c4}
\end{figure}

As shown in Fig.~\ref{fig:instrument_fft_c4}, even for the same pitch (\texttt{C4}), amplitude distributions around $f_0$ and harmonic components vary across instruments. Such differences likely influence the internal pitch representations learned by MFMs.

\subsection{Music Foundation Models}\label{sec:aboutMFM} 
We focus on two generative MFMs used in our experiments: Jukebox~\cite{jukebox} and MusicGen~\cite{musicgen}, both pretrained on large-scale music data via SSL.

\noindent\textbf{Jukebox.} Jukebox uses a vector quantized variational autoencoder~\cite{vq-vae} to convert music signals into discrete tokens, which are then autoregressively predicted by a hierarchical Transformer~\cite{transformer} decoder. The decoder consists of three levels: the top captures long-term musical features, while the middle and bottom refine temporal resolution.
% long-term audio features

\noindent\textbf{MusicGen.} MusicGen is a non-hierarchical generative model consisting of a pretrained neural audio codec~\cite{EnCodec}, a T5 text encoder~\cite{t5}, and a Transformer decoder.

\section{Method}\label{sec:method}
    \subsection{Extracting Pitch-Dependent Features}\label{sec:method_pitch_feature}
% As shown in Fig.~\ref{fig:concept}, we input monophonic music signals with specific pitches into an MFM and extract intermediate representations from each Transformer-decoder layer. Following prior work~\cite{Syntheory}, we obtain a time series of intermediate representations for each isolated musical note and average them over time to obtain a single vector representing the entire note. This process is repeated for $N_{\mathrm{key}}$ pitches across all layers, resulting in a set of pitch-conditioned intermediate representation vectors for each layer.

As shown in Fig.~\ref{fig:concept}, we input monophonic music signals with specific pitches into the model's pretrained audio tokenizer and pass the resulting token sequences to the Transformer decoder. Following prior work~\cite{Syntheory}, we obtain a time series of intermediate representations from each Transformer-decoder layer and average them over time to obtain one vector per isolated musical note. Repeating this process for $N_{\mathrm{key}}$ pitches yields a set of pitch-conditioned intermediate representation vectors for each layer.

To capture pitch-related structure, we apply principal component analysis (PCA) to the set of pitch-conditioned intermediate representations obtained from each layer. We compute the top five principal components and enumerate all $\binom{5}{3} = 10$ combinations of three components. For each combination, the pitch-conditioned representations are projected into a three-dimensional space, resulting in ten different sets of three-dimensional embeddings for each layer. These embeddings are then used to evaluate the degree to which each layer encodes a helical structure with respect to pitch.

% 3D → three-dimensional に統一
% initial height → height offset に統一 
% three-dimensional embeddings に統一
% c,u,vについて，もっと詳細説明する？

\subsection{Parametric Pitch Helix Model}\label{sec:method_helicality}
We model the three-dimensional embeddings using a parametric helical function defined by nine parameters: the height offset $h_{\mathrm{0}}$, height slope $h_{\mathrm{pitch}}$, radius offset $r_{\mathrm{0}}$, radius slope $r_{\mathrm{slope}}$, angular frequency $\omega_{\mathrm{chroma}}$, phase offset $p_{\mathrm{0}}$, and three orthonormal basis vectors $\bm{c}, \bm{u}, \bm{v} \in \mathbb{R}^3$. Here, $\bm{c}$ represents the central axis of the helix, while $\bm{u}$ and $\bm{v}$ span the rotation plane. These parameters control the position, shape, and pitch-wise rotation of the helix.

\begin{figure}[t]
  \centering
  \includegraphics[height=3.5cm]{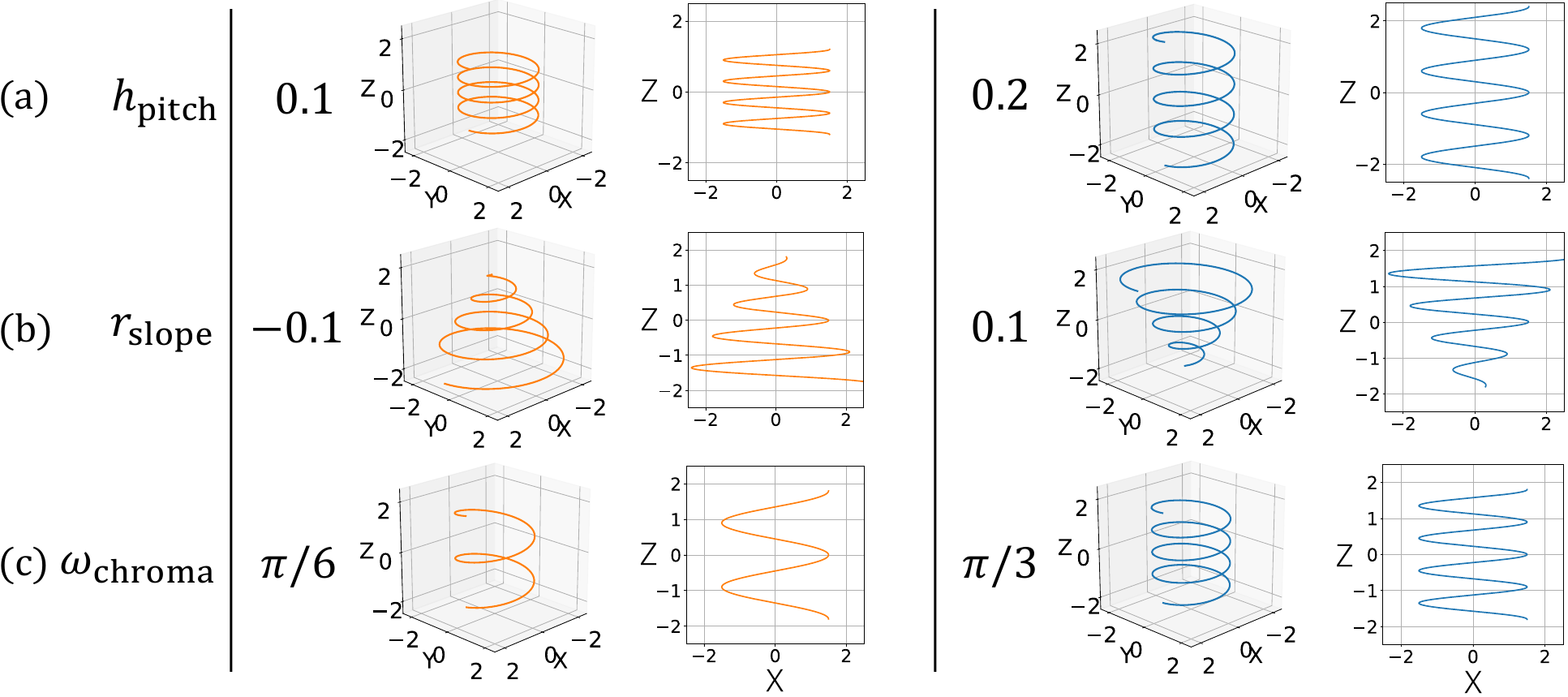}
  \vspace{-3mm}
  \caption{Effects of parameters in Eq.~\eqref{eq:helical_model}. (a) \( h_{\mathrm{pitch}} \) controls vertical spacing,  (b) \( r_{\mathrm{slope}} \) affects conicity (positive: expanding, negative: contracting),  (c) \( \omega_{\mathrm{chroma}} \) sets the number of pitches per rotation.  Each column shows a 3D view (left) and its \( xz \)-plane projection (right).
  }
  \label{fig:pitch_helix_model}
\end{figure}

Using these parameters, the helix function $\bm{y}(p)$ for the
semitone-ordered pitch index $p=1,\ldots,N_{\mathrm{key}}$ is given by:
\begin{align}
  \bm{y}(p) &= h(p) \cdot \bm{c} + r(p) \left( \cos\theta(p) \cdot \bm{u} + \sin\theta(p) \cdot \bm{v} \right) \notag \\
                &\text{where }
  \begin{cases}
    h(p) = h_{\mathrm{pitch}} \cdot p + h_{\mathrm{0}} \\
    r(p) = r_{\mathrm{slope}} \cdot p + r_{\mathrm{0}} \\
    \theta(p) = \omega_{\mathrm{chroma}} \cdot (p - p_{\mathrm{0}}).
  \end{cases}
  \label{eq:helical_model}
\end{align}
$h(p)$, $r(p)$, and $\theta(p)$ denote the height, radius, and phase, respectively. They are linear functions of $p$~(Fig.~\ref{fig:pitch_helix_model}). While the linear increase of $h(p)$ and $\theta(p)$ follows Shepard’s pitch helix~(Fig.~\ref{fig:pitch_helix}), $r(p)$ is also defined linearly to capture the cone-like variations observed in our experiments.
% (see Sec.~\ref{sec:slope}). 
% The parameters of $\bm{y}(t)$ are optimized to minimize the mean squared error (MSE) with respect to the three-dimensional embeddings.
The parameters of $\bm{y}(p)$ are optimized within predefined search ranges to minimize the mean squared error (MSE) with respect to the three-dimensional embeddings.

\subsection{Helicality} \label{sec:helicality}
 % score消す
To quantify how well the three-dimensional embeddings align with the helical model in Eq.~\eqref{eq:helical_model}, we apply the Helicality~\cite{helicality} as the inverse of the MSE. A higher score indicates a closer fit to the helical structure.

Formally, given three-dimensional embeddings $\{\bm{x}_p \}_{p=1}^{N_{\mathrm{key}}}$, the Helicality is defined as:
\begin{align}
\mathrm{Helicality}  = \left( \frac{1}{N_{\mathrm{key}}} \sum_{p=1}^{N_{\mathrm{key}}} \left| \bm{x}_p - \bm{y}(p) \right|^2 \right)^{-1}.
\label{eq:helical_score}
\end{align}
We compute this score for each of the $\binom{5}{3}$ projections described in Sec.~\ref{sec:method_pitch_feature}, and use the maximum score as the Helicality for each layer.

\subsection{Analysis Using Artificial Test Signals}\label{sec:method_harmonic_contribution}

% In this section, we analyze which harmonic components drive the clarity of the internal pitch helix, measured by Helicality. We synthesize artificial tones with controlled harmonic compositions, compute Helicality for each signal, and apply multiple regression to quantify contributing components.
% [takamichi]
% Section 2.4 で説明したように，楽器音に含まれる調波の有無あるいは強度によって音色が変わる．ここでは，人工的な test signal を作成して Helicality を図ることで，どの信号成分によって螺旋が形成されるかを調査する方法論を説明する．
As described in Sec.~\ref{sec:timbre}, the presence and amplitude of harmonic components in instrument sounds affect timbre. In this section, we describe a methodology for investigating which signal components contribute to helix formation by measuring Helicality with artificial test signals.

\subsubsection{Synthesis of Harmonic-Controlled Artificial Signals}

\begin{figure}[t]
  \centering
  \includegraphics[width=0.90\linewidth]{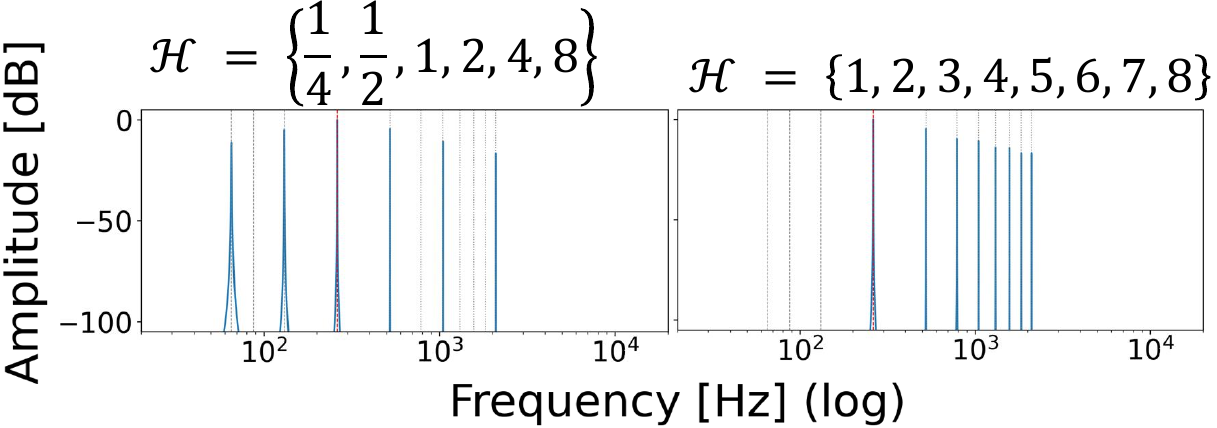}
  \vspace{-3mm}
  \caption{Amplitude spectra for different $\mathcal{H}$ with a fixed $f_0$ and $\mathrm{decay}$ tilt.}
  \label{fig:H_spectrum}
\end{figure}

An artificial signal $s_{\mathcal{H},f_0}^{(d)}(t)$ at time $t$ is defined as:
\begin{equation}
s_{\mathcal{H},f_0}^{(d)}(t)
= \sum_{h \in \mathcal{H}} a_h^{(d)}\sin\!\left(2\pi f_0 h t\right), 
\label{eq:harmonic_synthesis}
\end{equation}
where $\mathcal{H}$ denotes the set of harmonic indices included in the signal. The fundamental component $h=1$ is always included in $\mathcal{H}$, while integer harmonics $h=2,3,\ldots$ and subharmonics $h=\tfrac{1}{2},\,\tfrac{1}{3},\,\ldots$ are selectively included. Let $\mathcal{H}_{\mathrm{all}}$ denote the set of all possible harmonic indices, i.e., $\mathcal{H} \subset \mathcal{H}_{\mathrm{all}}$. The effects of different $\mathcal{H}$ are shown in Fig.~\ref{fig:H_spectrum}.

\begin{figure}[t]
  \centering
  \includegraphics[width=0.90\linewidth]{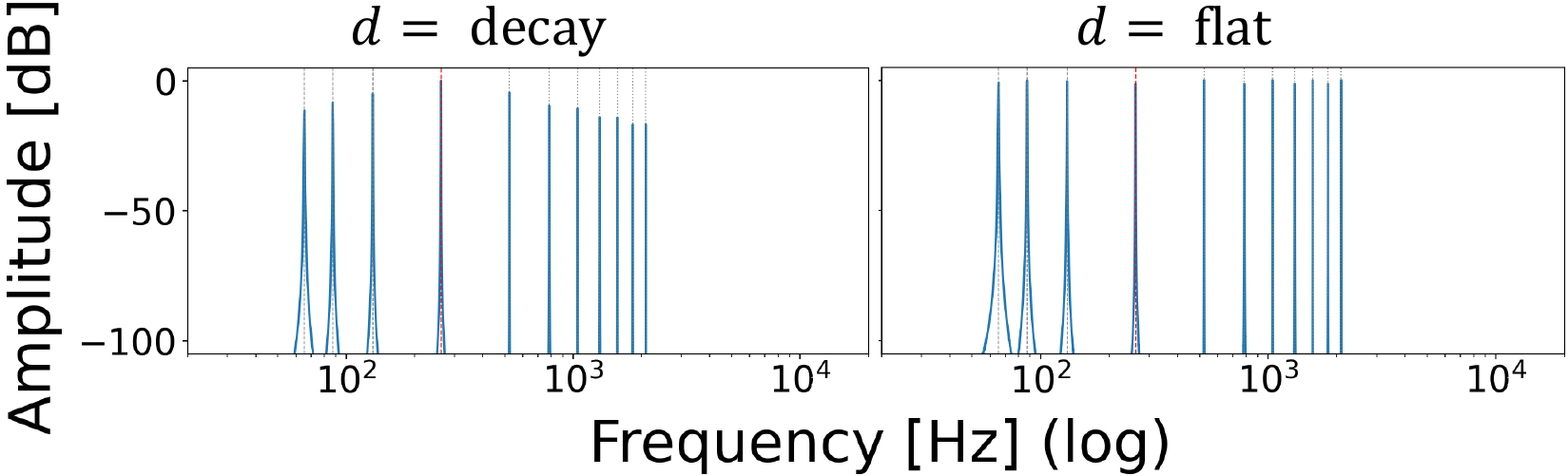}
  \vspace{-3mm}
  \caption{Amplitude spectra for different tilt conditions with a fixed $f_0$ and $\mathcal{H}$.}
  \label{fig:tilt}
\end{figure}

% A variable $d$ denotes the spectral tilt, with $\mathcal{D}=\{\mathrm{flat},\mathrm{decay}\}$.
A variable $d$ denotes the spectral tilt, where $d \in \mathcal{D} = \{\mathrm{flat}, \mathrm{decay}\}$.
To control the amplitude of each harmonic component, we define the amplitude coefficients $a_h^{(d)}$ as:
\begin{align}
a_h^{(\mathrm{flat})} &= 1 \label{eq:tilt_flat},\\
a_h^{(\mathrm{decay})} &= \min(h,\,1/h) \label{eq:tilt_decay}.
\end{align}
Equation~\eqref{eq:tilt_flat} corresponds to equal-amplitude harmonics, whereas Eq.~\eqref{eq:tilt_decay} applies symmetric amplitude decay as the frequency ratio deviates from $f_0$ (Fig.~\ref{fig:tilt}). These conditions yield different timbres for the same harmonic index $h$.

\subsubsection{Helicality for Each Condition}

For each harmonic condition $(\mathcal{H},d)$, we synthesize $N_{\mathrm{key}}$ signals with different pitches. We denote the set of the signals as
$\mathcal{S}_{\mathcal{H}}^{(d)} = \{s_{\mathcal{H},f_0}^{(d)}\}_{f_0 \in \mathcal{F}}$, 
where $\mathcal{F}$ is the set of fundamental frequencies corresponding to a predefined pitch range. The signals are input into a target MFM, and Helicality is computed for each layer as described in Sec.~\ref{sec:helicality}.

For the statistical analysis, we consider two dependent variables that summarize the layer-wise distribution of Helicality for each $\mathcal{S}_{\mathcal{H}}^{(d)}$. Let $H_{\ell}(\mathcal{S}_{\mathcal{H}}^{(d)})$ denote the Helicality at layer $\ell$. We then define the dependent variables as:
\begin{align}
y_{\mathrm{sum}}(\mathcal{S}_{\mathcal{H}}^{(d)}) &= \sum_{\mathrm{all}\;\ell} H_{\ell}(\mathcal{S}_{\mathcal{H}}^{(d)}) \label{eq:dv_sum},\\
y_{\mathrm{peak}}(\mathcal{S}_{\mathcal{H}}^{(d)}) &= \max_{\mathrm{all}\;\ell} H_{\ell}(\mathcal{S}_{\mathcal{H}}^{(d)}) \label{eq:dv_peak}.
\end{align}
We refer to $y_{\mathrm{sum}}$ as the \emph{layer-sum Helicality} and $y_{\mathrm{peak}}$ as the \emph{layer-max Helicality}. ``$\mathrm{all}\;\ell$'' means these calculations take into account all the Transformer layers. $y_{\mathrm{sum}}$ represents the overall strength of the internal pitch helix across all layers, while $y_{\mathrm{peak}}$ represents the strength of the most prominent layer.

\subsubsection{Multiple Regression Analysis and Statistical Testing}

We represent each harmonic condition $(\mathcal{H},d)$ using binary variables corresponding to combinations of harmonic indices and spectral tilt conditions, and analyze their relationship with Helicality via multiple regression. The binary variable is defined as:
\begin{equation}
x_{h,d} =
\begin{cases}
1, & \text{if } (h \in \mathcal{H} \land d' = d) \\
0, & \text{otherwise},
\end{cases}
\label{eq:dummy_harm_tilt}
\end{equation}
where $d'$ denotes the spectral tilt of the given condition.
% For example, $(h,l)=(2,\mathrm{flat})$, i.e., $x_{2,\mathrm{flat}}$, represents whether the second harmonic component is included under the flat spectral tilt condition.

For each dependent variable $y \in \{y_{\mathrm{sum}},\,y_{\mathrm{peak}}\}$, we assume the following linear model:
\begin{equation}
y
=
\beta_0
+
\sum_{(h,d)\in (\mathcal{H}_{\mathrm{all}}\setminus\{1\})\times\mathcal{D}}
\beta_{h,d}\,x_{h,d}
+
\varepsilon,
\label{eq:mlr}
\end{equation}
where $\beta_0$ is the intercept, $\beta_{h,d}$ is the regression coefficient representing the contribution of harmonic $h$ under the tilt condition $d$, $\times$ denotes the Cartesian product, and $\varepsilon$ is the error term. The fundamental component $h=1$ is excluded from the regression, so the summation is taken over $\mathcal{H}_{\mathrm{all}}\setminus\{1\}$.
We construct $(x_{h,d}, y)$ for all harmonic conditions and estimate the coefficients $\beta_{h,d}$ using ordinary least squares. We then perform an F-test to evaluate the significance of the overall regression model, followed by $t$-tests for each coefficient to assess whether the corresponding $(h,d)$ has a statistically significant effect on Helicality.

% Through this analysis, we identify (i) harmonic components that enhance the helical structure across layers via $y_{\mathrm{sum}}$, and (ii) harmonic components that contribute to peak helical structure in specific layers via $y_{\mathrm{peak}}$, with respect to harmonic index and spectral tilt.

\section{Experiments}\label{sec:experiments}
    \subsection{Experimental Setup} 
\subsubsection{Models} 
We analyze two trained MFMs: Jukebox and MusicGen.

\noindent\textbf{Jukebox.} 
We used the top-level decoder of the 5B model from \texttt{jukemirlib}~\cite{jukemir}\footnote{Available by appending a file name (e.g., \url{5b/vqvae.pth.tar}, \url{5b/prior_level_2.pth.tar}) to  \url{https://openaipublic.azureedge.net/jukebox/models/}.}.
The Transformer decoder consists of $72$ layers, and the dimensionality of the intermediate representations is $4,800$.

\noindent\textbf{MusicGen.} 
We used the large model (3.3B)\footnote{\url{https://huggingface.co/facebook/musicgen-large}}. 
Text conditioning was set to an empty string, and the input music signals were encoded using the frozen EnCodec. MusicGen's Transformer decoder has $48$ layers, and the dimensionality of the intermediate representations is $2,048$.

\subsubsection{Data} 
% We used a modified version of the SynTheory~\cite{Syntheory} dataset, focusing on isolated piano notes.  
% Twelve pitch classes spanning three octaves (\texttt{C3}–\texttt{B5}, $N_{\mathrm{key}} = 36$) were rendered as half notes at $120$ BPM using the \texttt{Acoustic Grand Piano} timbre with \texttt{TimGM6mb.sf2}~\cite{TimGM6mb}.
We used two types of datasets: (i) instrument note dataset and (ii) artificial test signal dataset.
For both datasets, $12$ pitch classes spanning $3$ octaves (\texttt{C3}–\texttt{B5}, $N_{\mathrm{key}} = 36$) were rendered as half notes at $120$ BPM.

\noindent\textbf{Instrument note dataset.}
We used the \texttt{notes} subset of the SynTheory~\cite{Syntheory} dataset, which contains isolated musical notes from $92$ instruments synthesized with \texttt{TimGM6mb.sf2}~\cite{TimGM6mb}, based on General MIDI~\cite{general_midi} program numbers and instrument categories.

\noindent\textbf{Artificial test signal dataset.}
We synthesized artificial signals based on Eq.~(\ref{eq:harmonic_synthesis}), in which harmonic components were explicitly controlled.  
% We defined multiple harmonic index sets $\mathcal{H}$ including integer harmonics ($h = 1,\dots,8$) and subharmonics ($h = \tfrac{1}{2}, \tfrac{1}{3}, \tfrac{1}{4}$), covering representative configurations of harmonic structures, including:
% all harmonics, odd-only, even-only, powers of two, single and paired harmonics, and sets including subharmonics.
We defined $\mathcal{H}_{\mathrm{all}}$ as $\{\tfrac{1}{4},\tfrac{1}{3},\tfrac{1}{2},1, \ldots,8\}$.
$\mathcal{H}$s were set to cover possible choices from $\mathcal{H}_{\mathrm{all}}$, and spectral tilt conditions $\mathcal{D} = \{\mathrm{flat}, \mathrm{decay}\}$ were applied to each $\mathcal{H}$, resulting in $137$ harmonic conditions in total.
% For each $\mathcal{H}$, we applied spectral tilt conditions $\mathcal{T} = \{\mathrm{flat}, \mathrm{decay}\}$, resulting in $137$ harmonic conditions in total.

\noindent\textbf{Preprocessing.}
Original signals were $4$-second stereo audio sampled at $44.1$~kHz, and they were converted to monaural. For MusicGen, $4$-second, $32$~kHz, monaural format was used. For Jukebox, the original signals were repeated or trimmed to be $24$ seconds.

\subsubsection{Parameter Optimization}\label{sec:fitting} 
We optimized the parameters of Eq.~(\ref{eq:helical_model}) using Optuna~\cite{optuna}.  
% For each set of three-dimensional embeddings, we ran $2000$ optimization trials, repeated three times with different random seeds.  
Each optimization used more than $1,000$ trials and was repeated three times with different random seeds.
These hyperparameters were chosen based on preliminary experiments to ensure that the MSE values sufficiently converged.  
The same search ranges were used for all layers, instruments, and models, e.g., the search range for $\omega_{\mathrm{chroma}}$ was set to 
$[-\frac{\pi}{2}, -\frac{\pi}{6}] \cup [\frac{\pi}{6}, \frac{\pi}{2}]$.

\subsection{Comparison among Instruments} \label{sec:inst_max_helicality} 

% To clarify the relationship between timbre and the internal pitch helix, we computed layer-wise maximum Helicality for each instrument included in the instrument note dataset for both Jukebox and MusicGen. The resulting scores showed a positive correlation between the two models (Pearson's $r = XX$); therefore, we present the Jukebox results in the following.

\begin{figure}[t]
  \centering
  \includegraphics[width=0.75\linewidth]{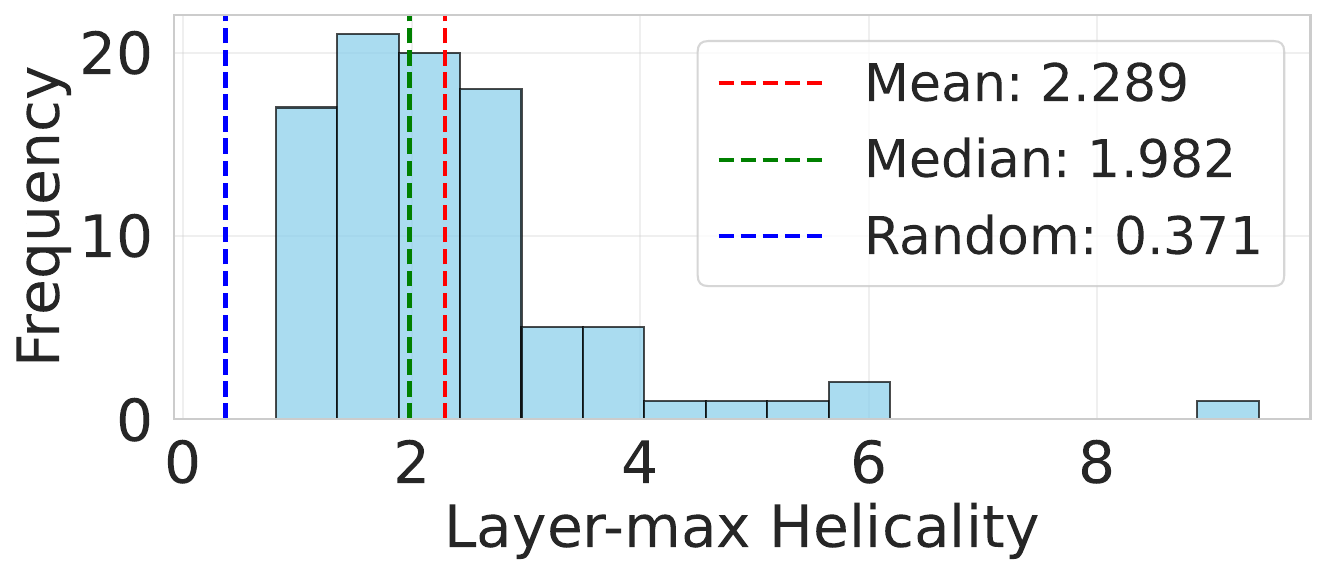}
  \vspace{-3mm}
  \caption{Distribution of layer-max Helicality across all instruments for Jukebox. The dashed lines indicate the mean, median, and the reference score.}
  \label{fig:max_helicality_hist}
\end{figure}

\begin{figure}[t]
  \centering
  \includegraphics[width=0.98\linewidth]{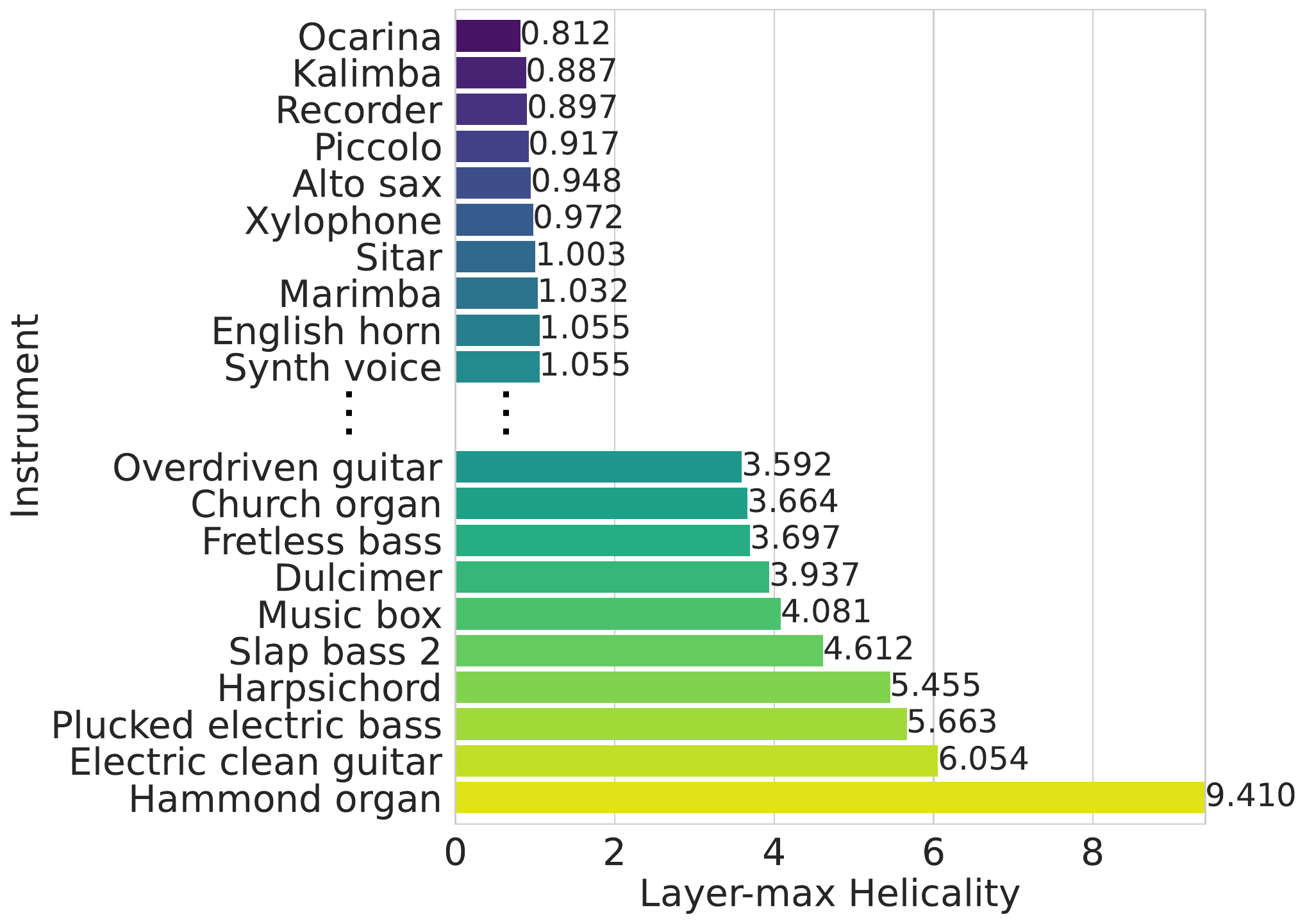}
  \vspace{-3mm}
  \caption{Best / worst 10 instruments by layer-max Helicality for Jukebox.}
  \label{fig:max_helicality_top_bottom15}
\end{figure}

\begin{figure}[t]
  \centering
  \includegraphics[width=0.94\linewidth]{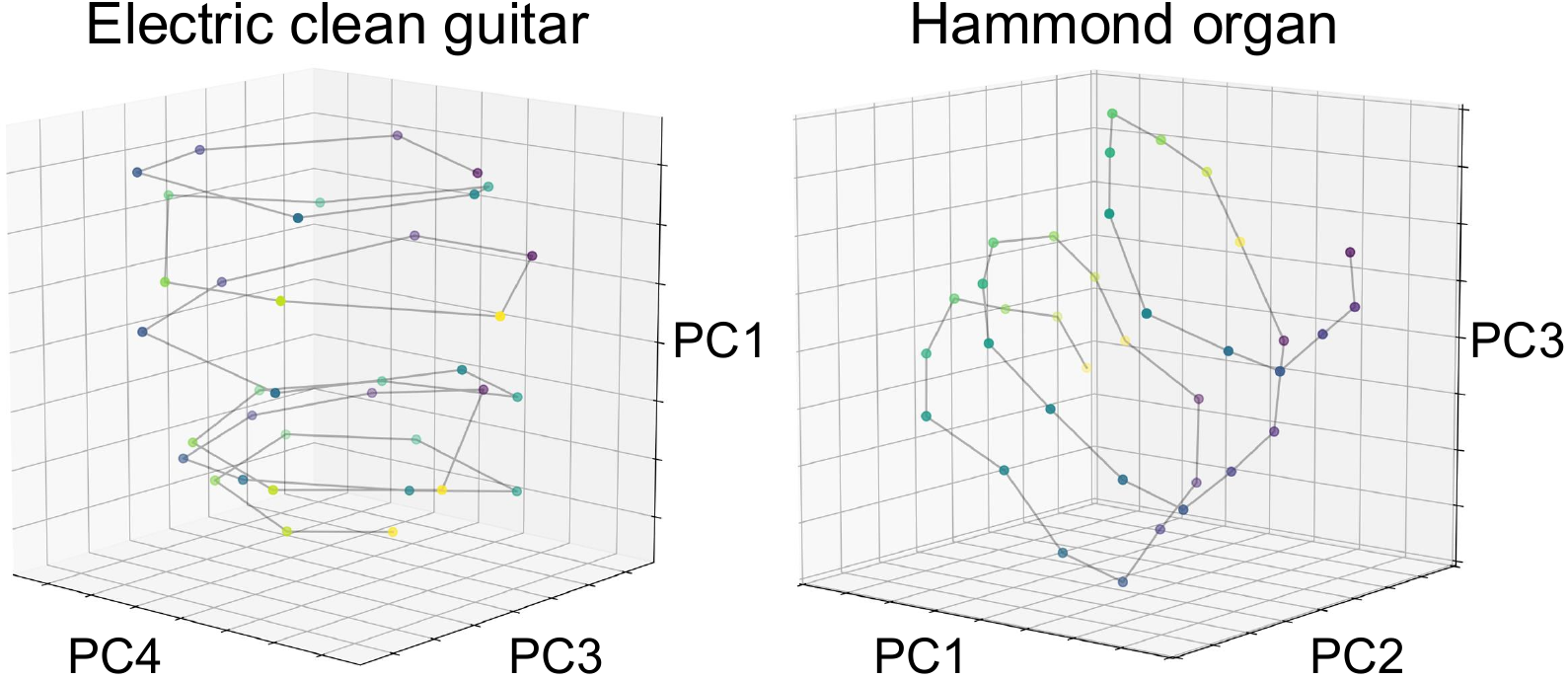}
  \vspace{-3mm}
  \caption{Jukebox embeddings for two instruments at the layer and three-PC projection corresponding to each instrument's layer-max Helicality.}
  \label{fig:ex_helix}
\end{figure}

\begin{figure}[t]
\centering
\includegraphics[width=0.88\linewidth]{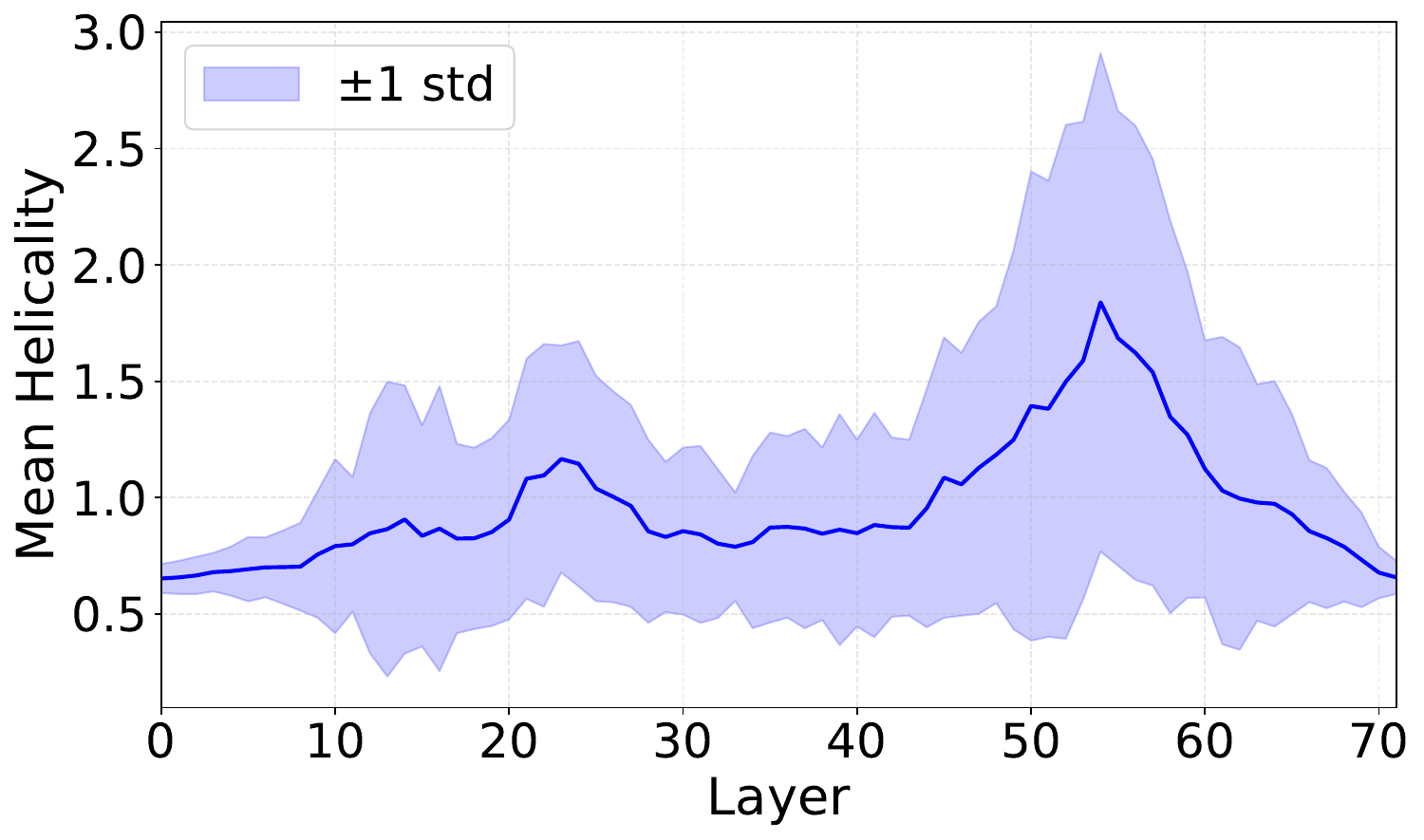}
\vspace{-3mm}
\caption{Mean Helicality across instruments at each layer of Jukebox.}
\label{fig:mean_heliclaity}
\end{figure}

\begin{figure}[t]
  \centering
  \includegraphics[width=0.98\linewidth]{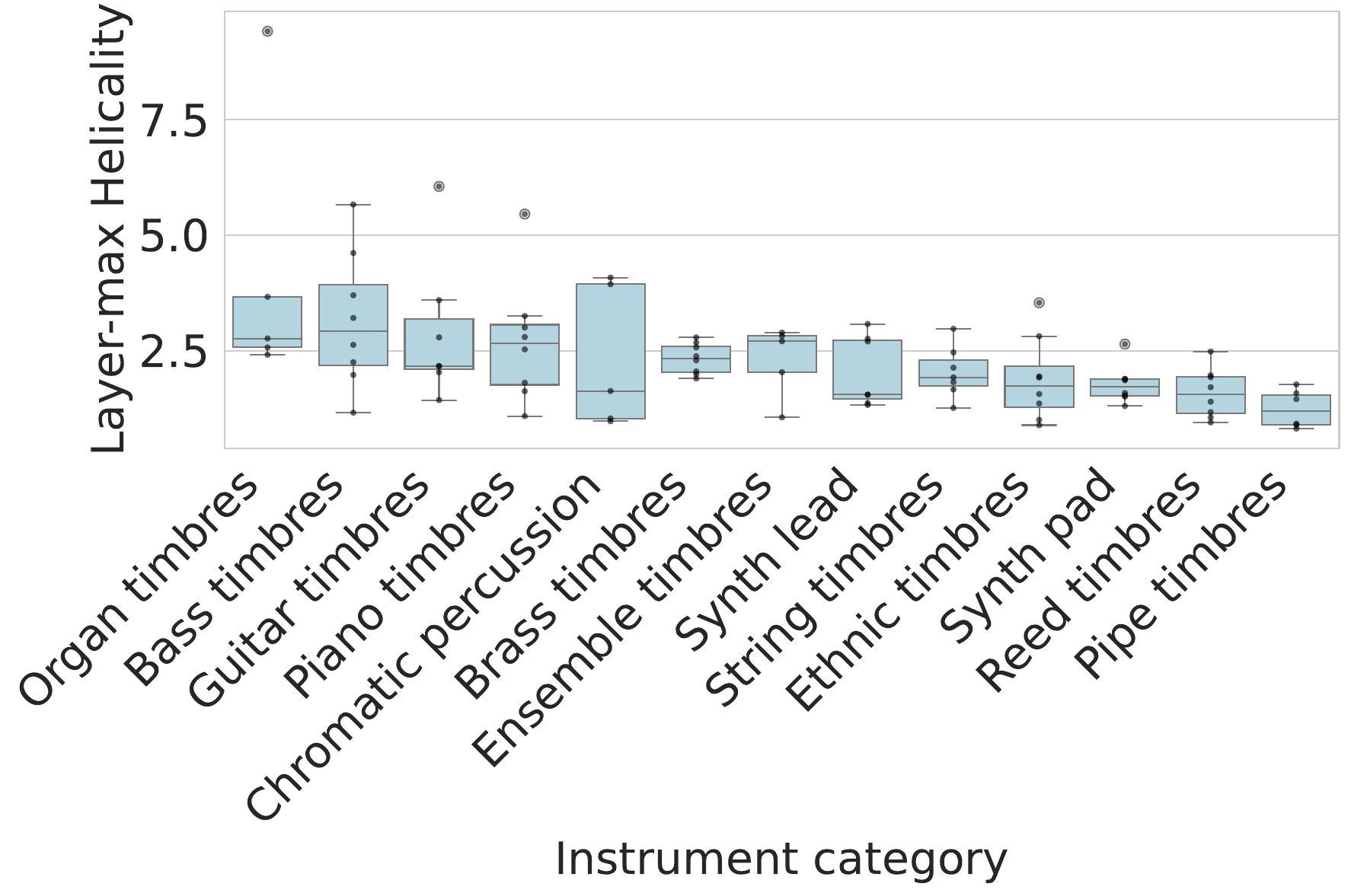}
  \vspace{-3mm}
  \caption{Distribution of layer-max Helicality across instrument categories for Jukebox. Each point represents the score of an individual instrument.}
  \label{fig:max_helicality_category}
\end{figure}

We investigated whether MFMs embed pitch in a helical structure, and if so, how the degree of Helicality varies across different instruments.
For each instrument, we computed Helicality at each layer from its signals, and defined the layer-max Helicality as the score for that instrument. In addition, we computed Helicality for three-dimensional uniform random variables (``Random'') as a reference.
We generated random variables $100$ times and computed their Helicality, obtaining a score of $0.371 \pm 0.014$. If the Helicality of a given instrument exceeds this value, we regard it as exhibiting a helical structure.

Figure~\ref{fig:max_helicality_hist} shows a histogram of Helicality, and Fig.~\ref{fig:max_helicality_top_bottom15} shows the best and worst $10$ instruments ranked by Helicality. The model used to obtain these results is Jukebox.

\textbf{MFMs embed pitch in a helical structure.}
As shown in Fig.~\ref{fig:max_helicality_hist}, the Helicality of all instruments exceeds that of ``Random''. Although not shown due to space limitations, similar results are observed for MusicGen. Therefore, we conclude that both analyzed MFMs embed pitch in a helical structure. However, the average Helicality across all instruments differs between models (Jukebox: $2.28$, MusicGen: $1.16$).

\textbf{The helical structure varies depending on instrument timbre.}
We focus on the differences across instruments. As shown in Fig.~\ref{fig:max_helicality_top_bottom15}, the scores vary widely, with a minimum of $0.812$ for ocarina and a maximum of $9.410$ for hammond organ. As shown in Fig.~\ref{fig:max_helicality_hist}, most instruments are distributed in a moderate range from $1$ to $3$, while a subset exhibits extremely high scores. Figure~\ref{fig:ex_helix} shows examples of helical structures for different instruments.
These examples illustrate that the helical structures differ across instruments, indicating that instrument timbre influences the geometry of the internal pitch helix. For instance, while the hammond organ exhibits a clear helix with one rotation per octave, the electric clean guitar shows a helix with two rotations per octave.
% Electric clean guitar exhibits a helical structure with two rotations per octave, whereas hammond organ exhibits a clearer helix with one rotation per octave, more closely reflecting octave equivalence.
% These results indicate that the structure of the helix is influenced by instrument timbre.
A similar trend is observed for MusicGen. Comparing the instrument rankings as illustrated in Fig.~\ref{fig:max_helicality_top_bottom15} between Jukebox and MusicGen, we obtained a linear correlation coefficient of $r = 0.59$. These results suggest that the relative ordering of Helicality across instruments tends to be consistent across different models.

\textbf{The prominent layer tends to appear in deeper layers.}
Figure~\ref{fig:mean_heliclaity} shows the mean Helicality across instruments at each layer. The mean Helicality peaks in deeper layers. These results suggest that abstract concepts such as pitch are more clearly represented in deeper layers. 
This tendency is consistent with prior probing studies~\cite{Syntheory, symbolic_concepts}, where deeper layers tend to perform better on downstream tasks. This suggests that Helicality may be related to downstream performance.

\textbf{Helicality shows trends across instrument categories.}
Figure~\ref{fig:max_helicality_category} shows the distribution of Helicality across instrument categories based on General MIDI. Organ timbres show the highest mean with large variance. Bass, guitar, and piano timbres exhibit moderate means with similarly high variability. In contrast, brass and string timbres show relatively stable distributions. These results suggest that Helicality is influenced by both global tendencies corresponding to timbral characteristics of instrument categories and instrument-specific acoustic factors.

\subsection{Analysis Using Artificial Test Signals} \label{sec:harmonic_regression} 

\begin{figure*}[t]
\centering
\includegraphics[width=0.92\linewidth]{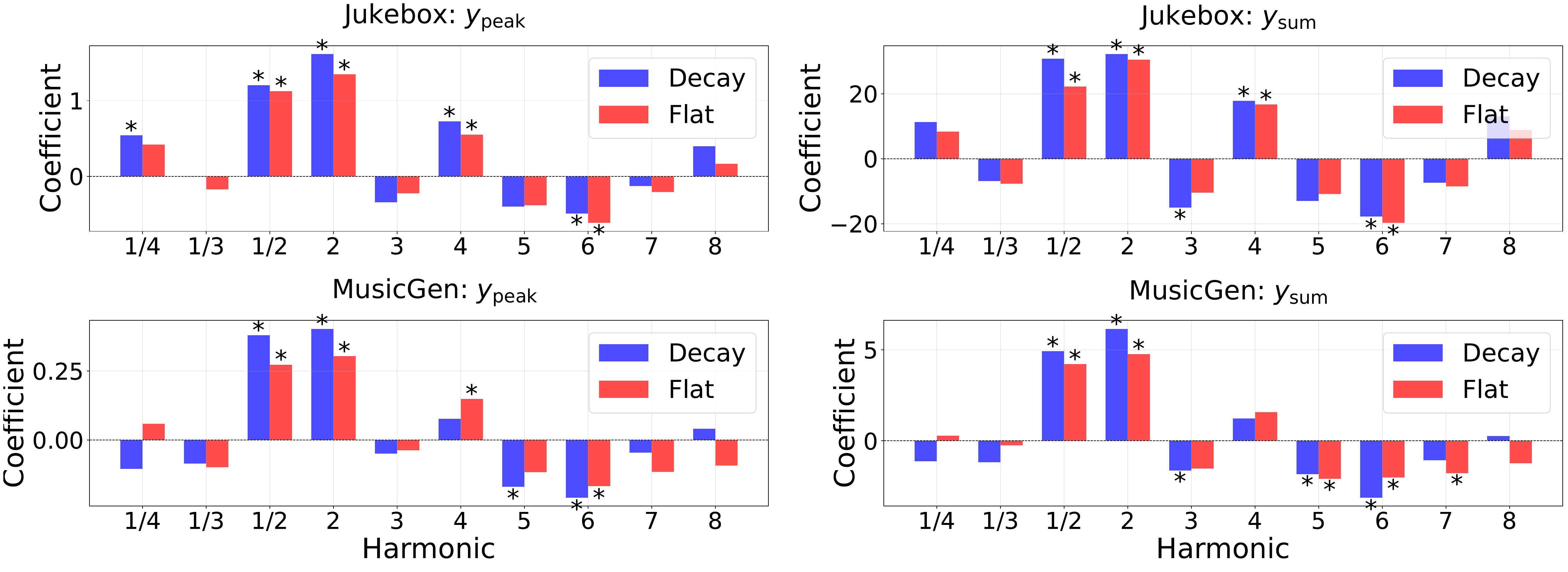}
\vspace{-3mm}
\caption{Regression coefficients for Jukebox and MusicGen. Asterisks indicate statistical significance ($p < 0.05$).}
\label{fig:reg_coef_combined_all}
\end{figure*}

\begin{table}[t]
  \centering
  \scriptsize
  \renewcommand{\arraystretch}{1.2}
  \begin{tabular}{c l l c c}
    \toprule
    Rank &
    Harmonic set $\mathcal{H}$ &
    Spectral tilt &
    Peak layer &
    $y_{\mathrm{peak}}$ \\
    \midrule
    1 &
    $\{ \tfrac{1}{2},\,1,\,2,\,4,\,8 \}$ &
    decay &
    52 &
    7.744 \\
    2 &
    $\{ \tfrac{1}{4},\tfrac{1}{2},\,1,\,2,\,4,\,8 \}$ &
    decay &
    51 &
    7.688 \\
    3 &
    $\{ \tfrac{1}{2},\,1,\,2 \}$ &
    flat &
    50 &
    6.906 \\
    4 &
    $\{ \tfrac{1}{2},\,1,\,2 \}$ &
    decay &
    56 &
    6.469 \\
    5 &
    $\{ \tfrac{1}{4},\tfrac{1}{2},\,1,\,2,\,4,\,8 \}$ &
    flat &
    53 &
    5.097 \\
    \midrule
    \vdots & \vdots & \vdots & \vdots & \vdots \\
    \midrule
    133 &
    $\{1\}$ (pure tone) &
    -- &
    21 &
    0.619 \\
    134 &
    $\{1,\,6,\,8\}$ &
    flat &
    36 &
    0.594 \\
    135 &
    $\{1,\,6,\,8\}$ &
    decay &
    1 &
    0.593 \\
    136 &
    $\{1,\,7\}$ &
    decay &
    0 &
    0.589 \\
    137 &
    $\{1,\,7,\,8\}$ &
    decay &
    0 &
    0.583 \\
    \bottomrule
  \end{tabular}
  \vspace{-1mm}
  \caption{Best and worst $5$ examples of layer-max Helicality for artificial test signals in Jukebox.}
  \label{tab:harmonic_helicality_examples}
\end{table}

% \begin{figure}
% \centering
% \includegraphics[width=0.98\linewidth]{figures/regression_coefficients_peak_helicality_MG.pdf}
% \vspace{-3mm}
% \caption{Regression coefficients for layer-max Helicality $y_{\mathrm{max}}$. Asterisks indicate statistical significance ($p < 0.05$).}
% \label{fig:reg_coef_max_mg}
% \end{figure}

% [takamichi] この順で書いて下さい
% In Sec.~\ref{sec:inst_max_helicality}, we investigated whether instrument timbre contributes to the formation of the internal pitch helix.また，Section X で説明したように，楽器の音色は harmonics の有無と強度で説明される．そこで，Based on the multiple regression analysis described in Sec.~\ref{sec:method_harmonic_contribution}, we analyzed whether the presence of each harmonic component and the spectral tilt contribute to the clarity of the internal pitch helix.目的変数は $y_{\mathrm{sum}}$ と $y_{\mathrm{peak}}$ とした．
% 図Xに．．．を表Xに．．．を示す．

In Sec.~\ref{sec:inst_max_helicality}, we investigated whether instrument timbre contributes to the formation of the internal pitch helix. As described in Section~\ref{sec:timbre}, instrument timbre can be characterized by the presence and amplitude of harmonic components. Based on the multiple regression analysis described in Sec.~\ref{sec:method_harmonic_contribution}, we analyzed whether the presence of each harmonic component and the spectral tilt contribute to the clarity of the internal pitch helix. The dependent variables were defined as $y_{\mathrm{sum}}$ and $y_{\mathrm{peak}}$.

Figure~\ref{fig:reg_coef_combined_all} shows the regression coefficients for $y_{\mathrm{sum}}$ and $y_{\mathrm{peak}}$ in both Jukebox and MusicGen. Table~\ref{tab:harmonic_helicality_examples} shows examples of layer-max Helicality obtained for each harmonic condition in Jukebox.
% As both figures exhibit similar trends, the following discussion applies to both.
The F-tests indicate that the regression models are statistically significant for both $y_{\mathrm{sum}}$ and $y_{\mathrm{peak}}$ in both models. Therefore, we interpret the contributions of each $(h, d)$ condition based on the corresponding $t$-tests of the regression coefficients.

% Building on this analysis, we used the artificial test signal dataset and computed the layer-wise maximum Helicality for each harmonic condition in both Jukebox and MusicGen. The results again showed a positive correlation between the models (Pearson's $r = 0.78$); therefore, as in Sec.~\ref{sec:inst_max_helicality}, we present the Jukebox results in the following.

% Based on the multiple regression analysis described in Sec.~\ref{sec:method_harmonic_contribution}, we analyzed whether the presence of each harmonic component and the spectral tilt contribute to the clarity of the internal pitch helix. 

% Fig.~\ref{fig:reg_coef_total} and Fig.~\ref{fig:reg_coef_peak} show the regression coefficients and corresponding $p$-values when using $y_{\mathrm{sum}}$ and $y_{\mathrm{peak}}$ as dependent variables, respectively. Since both figures exhibit similar trends, the following discussion applies to both.

\textbf{Octave-equivalent harmonics form the helix.}
As shown in Fig.~\ref{fig:reg_coef_combined_all}, the $\{2, 4, 8\}$-th harmonics and the $\{1/2, 1/4\}$ subharmonic components tend to exhibit positive coefficients, with $p$-values below $0.05$ in most cases. 
Consistently, as seen in Tab.~\ref{tab:harmonic_helicality_examples}, artificial signals with high Helicality are composed exclusively of harmonics whose frequency ratios are powers of $2$. A similar trend is observed for MusicGen, with a linear correlation of $r = 0.78$ between the models based on the Helicality rankings of artificial signals. These findings align with octave equivalence in human pitch perception and suggest that harmonic components with frequency ratios that are powers of $2$ play an important role in forming the internal pitch helix within MFMs.
Furthermore, this interpretation is supported by Sec.~\ref{sec:inst_max_helicality}: the hammond organ, which achieves the highest Helicality, exhibits a spectrum dominated by octave-equivalent harmonics, as shown in Fig.~\ref{fig:instrument_fft_c4}.

\textbf{Non-octave-equivalent harmonics hinder helix formation.}
In contrast, the remaining $\{3, 5, 6, 7\}$-th harmonics and the $1/3$ subharmonic component generally show negative coefficients. These results suggest that harmonics that are not octave-equivalent inhibit the formation of the helix. Indeed, the recorder, which has a low Helicality, exhibits a spectrum dominated by non-octave-equivalent harmonics, as shown in Fig.~\ref{fig:instrument_fft_c4}.

\textbf{Spectral decay facilitates helix formation.}
Focusing on the spectral tilt conditions, the decay condition tends to show larger coefficients than the flat condition for octave-equivalent harmonics. Compared to the flat condition with equal harmonic amplitudes, the decay condition, which resembles natural spectral decay, yields stronger contributions. This suggests that MFMs may form an internal pitch helix more effectively from spectral distributions that resemble those of natural instrument sounds.

\textbf{Signals without harmonics do not form a helix.}
As shown in Tab.~\ref{tab:harmonic_helicality_examples}, pure tone exhibits low Helicality. Moreover, the ocarina, which has the lowest Helicality, behaves as a single-mode Helmholtz resonator~\cite{physics_inst} and lacks a clear harmonic series. Therefore, it can be regarded as a condition similar to pure tone, leading to low Helicality. These observations indicate that signals without harmonic structure are less likely to form an internal pitch helix, which is consistent with the findings regarding the presence of harmonics described above.

\begin{comment}
\begin{figure}[t]
\centering
\includegraphics[width=0.9\linewidth]{figures/regression_coefficients_total_helicality.pdf}
\vspace{-3mm}
\caption{Regression coefficients for layer-sum Helicality $y_{\mathrm{sum}}$. Asterisks indicate statistical significance ($p < 0.05$).}
\label{fig:reg_coef_total}
\end{figure}

\begin{figure}[t]
\centering
\includegraphics[width=0.9\linewidth]{figures/regression_coefficients_peak_helicality.pdf}
\vspace{-3mm}
\caption{Regression coefficients for layer-max Helicality $y_{\mathrm{peak}}$.}
\label{fig:reg_coef_peak}
\end{figure}
\end{comment}

\section{Conclusion}
    This study demonstrated that the intermediate representations of two generative MFMs embed pitch information in a helical structure.
The clarity and geometry of this structure vary not only across the two analyzed models but also with the acoustic properties of the input, particularly its harmonic composition.
% Octave-equivalent harmonics promote the formation of the helix, while non-octave components hinder it, and natural spectral decay further enhances the structure. 
These results suggest that MFMs represent pitch not merely as frequency differences, but as a geometric structure grounded in harmonic relationships.
% These findings provide insight into how MFMs represent pitch and offer a foundation for improving the controllability of music generation models.
% However, further investigation of model generality, the influence of training data, and the relationship between Helicality and downstream task performance remains an important direction for future work.
However, whether the helical structure originates in the Transformer or earlier representations remains unclear. Further work should also examine its generality across models and training data, as well as its relationship to downstream task performance.

\clearpage  
\section{Acknowledgements}
This work was supported by JST FOREST Grant Number JPMJFR226V and JSPS KAKENHI Grant Number JP26K21256.

% For BibTeX users:
\bibliography{ISMIRtemplate}

% For non BibTeX users:
%\begin{thebibliography}{citations}
% \bibitem{Author:17}
% E.~Author and B.~Authour, ``The title of the conference paper,'' in {\em Proc.
% of the Int. Society for Music Information Retrieval Conf.}, (Suzhou, China),
% pp.~111--117, 2017.
%
% \bibitem{Someone:10}
% A.~Someone, B.~Someone, and C.~Someone, ``The title of the journal paper,''
%  {\em Journal of New Music Research}, vol.~A, pp.~111--222, September 2010.
%
% \bibitem{Person:20}
% O.~Person, {\em Title of the Book}.
% \newblock Montr\'{e}al, Canada: McGill-Queen's University Press, 2021.
%
% \bibitem{Person:09}
% F.~Person and S.~Person, ``Title of a chapter this book,'' in {\em A Book
% Containing Delightful Chapters} (A.~G. Editor, ed.), pp.~58--102, Tokyo,
% Japan: The Publisher, 2009.
%
%\end{thebibliography}

\end{document}